# Opinion Formation and the Collective Dynamics of Risk Perception

Mehdi Moussaïd[1,2]*

1 Center for Adaptive Behavior and Cognition, Max Planck Institute for Human Development, Berlin, Germany, 2 Center for Adaptive Rationality, Max Planck Institute for Human Development, Berlin, Germany

## Abstract

The formation of collective opinion is a complex phenomenon that results from the combined effects of mass media exposure and social influence between individuals. The present work introduces a model of opinion formation specifically designed to address risk judgments, such as attitudes towards climate change, terrorist threats, or children vaccination. The model assumes that people collect risk information from the media environment and exchange them locally with other individuals. Even though individuals are initially exposed to the same sample of information, the model predicts the emergence of opinion polarization and clustering. In particular, numerical simulations highlight two crucial factors that determine the collective outcome: the propensity of individuals to search for independent information, and the strength of social influence. This work provides a quantitative framework to anticipate and manage how the public responds to a given risk, and could help understanding the systemic amplification of fears and worries, or the underestimation of real dangers.

**Funding:** The author is funded by the Max Planck Society. The funders had no role in study design, data collection and analysis, decision to publish, or preparation of the manuscript.

**Competing Interests:** The author has declared that no competing interests exist.

* E-mail: mehdi.moussaid@gmail.com

## Introduction

With the ongoing growth of mass media and communication technologies, people are constantly exposed to steady flows of news information and subjective opinions of others about political ideas, emerging technologies, commercial products, or health-related threats. Pieces of information are broadcasted in mass media such as television, newspapers, or online recommendation systems, and further exchanged among individuals during personal conversations and through social networking tools such as Twitter or Facebook. As a result, people often need to integrate a large amount of conflicting and sometimes distorted information and peer opinions to form their own judgment on various social issues.

The question of how people form and revise opinions under the influence of others is at the heart of the field of opinion dynamics [1,2], which has been particularly active in the last decade [3–6]. In particular, existing research has demonstrated that local interactions among neighboring people often give rise to complex collective patterns of opinion formation [7–11]. Examples of such collective patterns are *consensus formation* where repeated local influences among people support the emergence of a global agreement in the population, *polarization* in situations where radically opposed opinions emerge and coexist the population [12–14], and *clustering* when local groups of like-minded people form simultaneously [15,16].

Nowadays, the study of social influence and opinion dynamics is becoming a central issue in modern societies. In fact, the easy access to media and social information exacerbates individuals exposition to news articles and peer opinions, which increasingly shapes their judgments in various domains, such as marketing [17,18], political science [19,20], or risk perception [21,22].

The present work specifically addresses the subject of opinion dynamics in the field of risk perception, in which individuals form and revise judgments about the possible danger of a hazardous activity or technology [23,24]. The research on risk perception aims at understanding, anticipating and managing how the public responds to a given risk or health issue, such as global warming [25], nanotechnologies [26], or vaccination [27]. While most research has focused on the social and psychological factors that influence the way an individual evaluates the severity of a given hazard, the *collective dynamics* of the system remains largely unexplored: What kind of collective patterns of risk perception emerge at the population level, and what are the underlying mechanisms of the system?

Similarly to other domains where opinion dynamics applies, the perception of risk exhibits typical signatures of self-organizing processes: First, individual risk judgments tend to be correlated with the proximity of individuals in their social network, suggesting a possibly significant inter-individual influence [28–30]. Furthermore, risk judgments are often polarized, with people expressing very high and very low levels of worries coexisting in the same population (see, e.g., the recent survey on food-related risks [31]). In particular, inter-individual discussions on a given hazard tend to support the amplification or attenuation of the individuals' risk perception [32], which is a common mechanism of group polarization [13]. As suggested by the social amplification of risk framework [21], various communication channels through which risk information flows play the role of "amplification stations" by transmitting a small and often biased subset of the available information. Such amplification stations could be the individuals themselves, or any channel of information such as public media. This suggests the existence of feedback loops and information cascade [33], whereby biased information would tend to become even more biased as it flows from one individual to another, leading to the amplification of the perception of certain risks, or reversely causing the consensual underestimation of real threats.

One important element that is currently missing in existing research is a proper simulation model that would generate concrete predictions about the collective patterns of risk judgments that emerge in a large population. While the social amplification of risk framework is very informative of the general processes that are involved in the system's dynamics, it remains too conceptual for conducting multi-agent simulations. Therefore, a more detailed model is needed to explore and understand quantitatively how macroscopic patterns of risk perception emerge at the population level.

In this work, an individual-based model of risk perception is introduced. The model draws upon existing models of opinion dynamics on the one hand, and empirical and theoretical concepts of risk communication on the other hand. For this, I provide a quantitative description of (1) how risk information propagates from the media to the individuals, (2) how individuals' form and revise their judgment based on the information they have received, and (3) how people communicate about the risk with others. In particular, the model assumes a cognitive bias whereby individuals integrate and communicate information in accordance with their current views [14]. I show that this bias is at the origin of a complex collective dynamics characterized by the emergence of polarized clusters of people having opposed risk perception, even though individuals are initially exposed to the same sample of information. Furthermore, the model allows drawing connections between aggregate search patterns observed over the Web (e.g. [34]), the average individual knowledge about the suspected risk, and the internal dynamics of risk perception. In particular, I show that the collective dynamics of the system is determined by two crucial factors: how much people search for their own independent information and how much they exchange information with their peers.

## The Model

We assume a media environment made of $N_{info}$ pieces of information, such as newspaper articles, or Web pages dealing with a particular risk. Each piece of information $k$ is characterized by a certain level of danger $D_k$ and a certain level of safety $S_k = 1 - D_k$, which describe how much the item emphasizes the danger or the safety of the situation, respectively. For instance, a piece of information with $\{D_k = 1, S_k = 0\}$ would be an alarmist item, whereas $\{D_k = 0, S_k = 1\}$ would be a reassuring item, and $\{D_k = 0.5, S_k = 0.5\}$ a well-balanced item. For the sake of simplicity, we assume here a simple and perfectly balanced distribution of information made of $N_{info} = 101$ items with $D_k$ ranging from 0 to 1 by step 0.01, formally defined as:

$$D_k = \frac{k-1}{N_{info} - 1} \quad (1)$$

for $k$ varying from 1 to $N_{info}$.

In this environment, $N$ individuals located over a square lattice of size L×L with periodic boundary conditions collect pieces of information from the media and exchange them with their neighbors. At each moment of time, the risk perception $r_i$ of an individual $i$ is derived from the list of items the individual owns (**Figure 1**).

Agents are additionally characterized by an awareness level $A_i$ describing how active they are in searching for information and discussing the issue with their friends [29]. The awareness level is assumed to increase by an offset $\delta = 1$ when the individual receives a novel piece of information but tends to gradually fade away at the speed of $\delta/2$ at every time step. In such a way, individuals actively search for information and discuss with their friends as they receive new information, but tend to loose interest in the issue otherwise. In the following, I describe the four steps of the elaboration of the model, which are depicted in **Figure 1**. In addition, **Table 1** and **Table 2** provide a summary of the parameters and variables that are used in the model.

### Media Influence (Step 1)

At each time step, individuals have a probability $P_{ind}$ to start searching for new information in the media, such as exploring the Web for news articles about the suspected risk. When they decide to do so, individuals discover one piece of information at random among the $N_{info}$ available in the environment. The probability $P_{ind}$ to start an independent search is given by

$$P_{ind} = \bar{A}_i \omega_{ind} + \varepsilon \quad (2)$$

where $\bar{A}_i = 1$ if the awareness of the individual is positive $A_i > 0$, and $\bar{A}_i = 0$ otherwise. The parameter $\omega_{ind}$ represents the tendency of people to search for their own independent information. Here, $\varepsilon$ is a small random value chosen in the interval $[0 \ 10^{-2}]$, such that individuals still have a small probability $\varepsilon$ to discover a piece of information by chance, even when their awareness is zero.

### Social Influence (Step 2)

In addition to their independent search behavior, individuals can also acquire pieces of information from their neighbors. At each time step, each individual has a probability $P_{social}$ to start a conversation with one random neighbor:

$$P_{social} = \bar{A}_i \omega_{social} + \varepsilon \quad (3)$$

where $\omega_{social}$ represents the tendency of people to discuss the issue with their neighbors, and $\bar{A}_i$ and $\varepsilon$ are defined as previously. When a conversation starts between an individual and a neighbor, both individuals select a piece of information among those they know about and communicate it to the other person. The piece of information that is communicated is the one that has the higher weight (see step 3 below). If several items have equal weights, the individual selects one at random among them. The weight of a piece of information is defined in the next step.

### Integration of a New Piece of Information (Step 3)

When a new piece of information $k$ is discovered (through social interactions or after an independent search), the individual gives it a weight $\theta_k^i$. The weight can represent various aspects of the information, such as the perceived credibility of the source, the novelty of the information [35], or how much it agrees with the individual's current view [26,36]. For the sake of simplicity, however, only the later factor is taken into account in the present model. In line with existing research in the field of opinion dynamics, we use a simple step function defined as follows:

$$\theta_k^i = 1, \text{ if } |r_i - D_k| < \tau \text{ (strong agreement)},$$
$$\theta_k^i = 0, \text{ if } |r_i - D_k| > 1 - \tau \text{ (strong disagreement)}, \quad (4)$$
$$\theta_k^i = \theta_0, \text{ otherwise}.$$

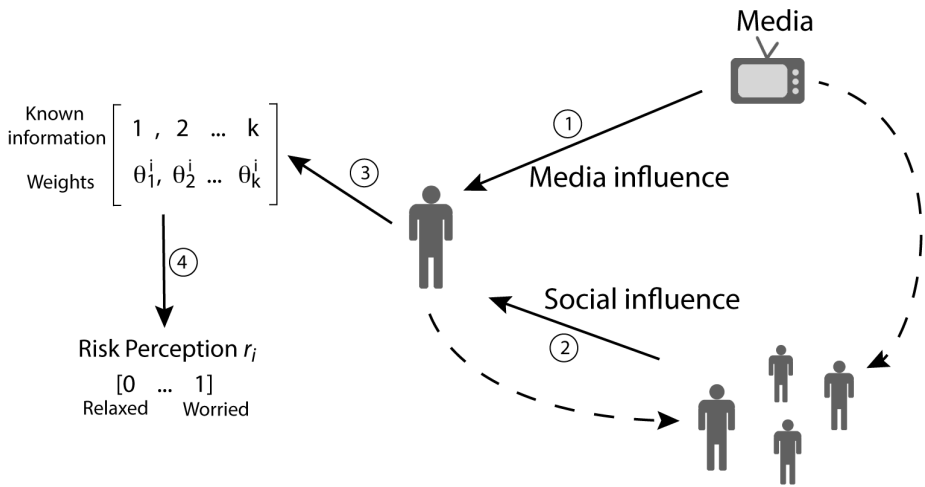

**Figure 1. Schematic representation of the model.** Individuals receive pieces of information from the media (1) and from their peers (2). Each piece of information $k$ is given a weight $\theta_k^i$ by the individual $i$ and stored in his or her memory (3). The collection of weighted information an individual $i$ owns is finally used to determine the level of risk perception $r_i$ (4). Circled numbers indicate different steps of the elaboration of the model as described in the main text. All model parameters and variables are summarized in tables 1 and 2, respectively.
doi:10.1371/journal.pone.0084592.g001

Where $\tau$ is a threshold value and $\theta_0$ is a model parameter. In such a way, the individual gives a strong consideration to pieces of information that agree with his or her current view, ignores those that strongly disagree with his or her current view, and moderately takes into account those that are at intermediate distance. This modeling approach is based on the Bounded Confidence model of opinion dynamics [7,37], the mechanisms of which have been confirmed under experimental conditions [38]. For the scope of this paper, we used $\tau = 0.2$ and $\theta_0 = 0.5$.

### The Construction of Risk Perception (Step 4)

Finally, the risk perception $r_i$ of an individual $i$ results from the integration of all pieces of information $k$ the individual owns, and their respective weights $\theta_k^i$. The integration function should obey some constraints. For example, an individual who has no information at all should have a risk perception $r_i = 0$ (i.e. the individual is not aware of any possible danger), whereas a large amount of well-balanced information should result in a risk perception $r_i = 0.5$ (i.e. the individual is unsure of the danger). In order to account for the above constraints, the risk integration is given by the equation:

$$f(\overline{S_i},\overline{D_i}) = 1 - \exp\left(-\frac{\alpha \overline{D_i}}{\overline{S_i} + \beta}\right) \quad (5)$$

where $\alpha$ and $\beta$ are model parameters. The variables $\overline{S_i} = \sum \theta_k^i S_k$, and $\overline{D_i} = \sum \theta_k^i D_k$ represent the weighted sum of the danger level

**Table 1.** Description of model parameters and the corresponding values used in the numerical simulations.

| Name | Description | Used value |
|---|---|---|
| *Social and media environment* | | |
| $N_{info}$ | Number of pieces of information available in the environment | 101 |
| $D_k, S_k$ | Level of danger (resp. safety) of each piece of information $k$, defined in the interval [0 1]. The two parameters are connected by the relation $D_k = 1 - S_k$. | Uniform distribution, see Eq. (1) |
| $N$ | Number of individuals in the population | 2500 |
| $\delta$ | Offset of the awareness level | 1 |
| *How people collect information in their environment* | | |
| $\omega_{ind}$ | Tendency to search for own information | $\omega_{ind} = 0.1$, in figure 6. |
| $\omega_{social}$ | Tendency to interact with other individuals | $\omega_{social} = 0.9$, in figure 6. |
| $\varepsilon$ | Noise parameter | Random value in [0 $10^{-2}$] |
| *How people integrate new pieces of information* | | |
| $\tau$ | Threshold value for the social influence model | $\tau = 0.2$ |
| $\theta_0$ | Default weight | $\theta_0 = 0.5$ |
| *How people construct and revise risk perception* | | |
| $\alpha, \beta$ | Parameters of the risk perception model (see Eq. 5 and Fig. 2) | $\alpha = 0.8$; $\beta = 0.2$; |

doi:10.1371/journal.pone.0084592.t001

**Table 2.** Description of model variables and their initial values as used in the numerical simulations.

| Name | Description | Value |
| --- | --- | --- |
| $r_i$ | Risk perception of individual $i$, defined in the interval [0 1]. | Initially set to $r_i = 0$. Then given by equation (5) |
| $A_i$ | Awareness level of individual $i$, defined in the interval $[0 +\infty]$. | Initially set to $A_i = 0$ |
| $P_{ind}$ | Probability to search for new information in the media. | Given by equation (2) |
| $P_{soc}$ | Probability to interact with other people. | Given by equation (3) |
| $\theta_k^i$ | Weight given to information $k$ by individual $i$ | Given by equation (4) |
| $\overline{S_i}, \overline{D_i}$ | Weighted sum of danger levels (resp. safety levels) of all pieces of information known by individual $i$. | Given by equation (5) |

doi:10.1371/journal.pone.0084592.t002

$D_k$ and the safety level $S_k$ of all the pieces of information $k$ known by the individual $i$. The figure 2 shows the shape of the function for parameters $\alpha = 0.8$, and $\beta = 0.2$, which will be used in the present work.

## Results

The predictions of the above model are now explored by means of computer simulations. The initial conditions of the simulations are set to N = 2500 agents, located over a square lattice of size 50×50, with the initial risk perception $r_i = 0$ for all individuals $i$. At each time step, agents simultaneously search for new information with a probability $P_{ind}$ and then exchange them with other agents in their Moore neighborhood (i.e. the eight individuals surrounding their own position) with a probability $P_{social}$. The simulation runs until the system has reached a stable state, i.e. after 500 runs.

First, I study specifically how the balance between independent search and social influence affect the collective dynamics of the system. For this, the two key parameters $\omega_{ind}$ and $\omega_{social}$ are gradually varied from 0 to 1 (see the modeling steps 1 and 2). As shown in figure 3, a rich variety of collective patterns can emerge depending on the combined values of $\omega_{ind}$ and $\omega_{social}$. When both parameter values are low, people occasionally discover some pieces of information in the media, which directly determine their risk judgment. Consequently, some individuals exhibit a high level of worry while others have a low risk perception, depending on the nature of information they discovered in the first place (Fig. 3a). As the strength of social influence increases, however, a strong correlation between neighboring people sets up (Fig. 3b). Even though agents are initially exposed to the same set of information, influences among neighboring people generate local amplification loops giving rise to the clustering pattern. Finally, a strong weight for the independent search parameter $\omega_{ind}$ generates a consensual risk perception in the population (Fig. 3c).

More specifically, the polarization of opinions can be simply measured as the standard deviation of the opinion distribution over the entire population. In such a way, the polarization is high if different opinions coexist in the population, whereas a global consensual judgment generates a low polarization value. As shown in figure 4a, the perception of risk tends to become homogeneous as the strength of independent search increases above $\omega_{ind} \approx 0.5$ (low polarization, in blue). This effect is due to the fact that individuals collecting a large amount of independent information will eventually end up with a similar knowledge of the problem and therefore develop a common risk judgment. In fact, the low polarization region visible in the upper part of the figure 4a also coincides with a parameter space where the agents are very well informed, as shown Figure 4d.

The polarization level alone, however, does not characterize the clustering level of the population well. For instance, the examples shown Figure 3a and 3b both exhibit a high polarization level (i.e. opposed opinions coexist in the population), but only the pattern in figure 3b displays clusters (i.e. local agreement between neighboring agents). Therefore, a local disagreement coefficient $D_i$ is introduced, which is defined as the average absolute difference between the opinion of an individual $i$ and the opinion of his or her direct neighbors $k$:

$$D_i = \sum |r_i - r_k| / N_k \qquad (6)$$

where $N_k = 8$ is the number of direct neighbors of agent $i$. Therefore, $D_i$ is low when the individual $i$ agrees with his or her neighbors, and $D_i$ is high in case of a strong local difference of opinions. The average value of $D_i$ over all individuals $i$ is shown figure 4b. Finally, the clustering level $C$ is obtained by dividing the polarization level by the average $D_i$, which yields a high clustering value when the opinions are polarized *and* when neighboring people having similar views. As shown in figure 4c, the clustering occurs only at the bottom right corner of the map, i.e. when social influence is strong and the tendency of independent search is low. Interestingly, this zone also corresponds to a region of the parameter space where people are less informed (Fig. 4d).

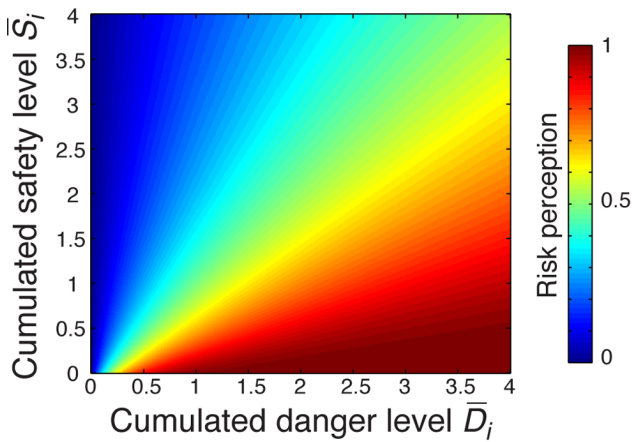

**Figure 2. Graphical representation of the risk perception function $f(\overline{S_i}, \overline{D_i})$.** The function indicates the perception of risk of an individual owning a total amount of information $\overline{D_i}$ and $\overline{S_i}$ indicating the danger and the safety of the situation, respectively. The function parameters are set to $\alpha = 0.8$, and $\beta = 0.2$. In the absence of any information, the risk perception level is 0, whereas large and well-balanced amounts of information for both sides yield a risk level of 0.5. The function always returns a value between 0 and 1.
doi:10.1371/journal.pone.0084592.g002

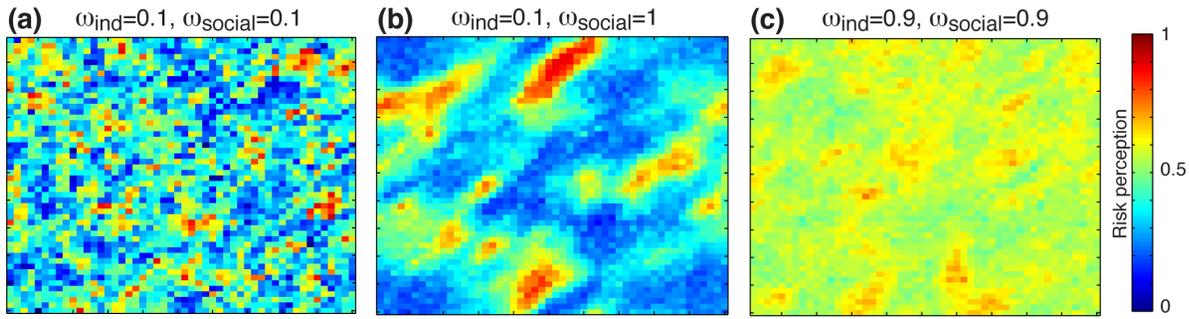

**Figure 3. Three representative examples of the model predictions.** (**a**) With low levels of independent search $\omega_{ind}$ and social influence $\omega_{social}$, opposed judgments coexist in the population but the clustering level is low. (**b**) As the weight of social influence increases, clusters of neighboring people with similar views emerge. (**c**) When the levels of independent search and social influence are both high, individuals tend to converge towards a global consensus with a risk perception close to 0.5, corresponding to a well-balance judgment. Simulations were conducted with N = 2500 agents (i.e. grid size of 50×50).
doi:10.1371/journal.pone.0084592.g003

Another property of the system is the activity patterns that emerge through local interactions among the agents. The figure 5 shows time series of the global search volume in the population, which is defined as the number of individuals per unit of time who engaged in an independent search for information. By varying the main parameters $\omega_{ind}$ and $\omega_{social}$, different patterns can be generated: a constant low search volume when both parameters are low (Fig. 5a), a spiky pattern followed by a relatively rapid relaxation (Fig. 5b), or a step-like pattern characterized by a period of intense activity followed by a sudden drop of the collective attention (Fig. 5c). This variety of outcomes results from two opposed mechanisms: On the one hand, agents are increasingly more likely to search and communicate about the risk issue as they receive new information because their awareness level increases, which generates the initial amplification of the activity. On the other hand, however, undiscovered pieces of information tend to become scarcer over time, which causes a decrease of the awareness level, resulting in the relaxation of the search pattern after a certain time.

The above results demonstrate the interesting flexibility of the model, and its ability to generate a rich variety of collective patterns. Is it unclear, however, what parameter values would better fit real life phenomena. Could we infer the most appropriate parameter values for $\omega_{ind}$ and $\omega_{social}$ by comparing the model predictions to existing empirical facts? First, it is known that risk perception is strongly polarized, as it has been shown in empirical risk surveys, for instance when asking people to evaluate the severity of various food-related risks [31], or during experimental studies [25].Therefore, the weight of independent search $\omega_{ind}$ is likely to have a low value (see figure 4a). Furthermore, recent social network analyses have highlighted the existence of opinion clustering, showing that individual risk judgments are correlated with the strength of the social ties between people [28]. With regard to the present model predictions, this suggests that the weight of social influence is strong, and that real life phenomena occur mostly around the bottom right corner of the maps presented in figure 4. Besides, this region of the parameter space is also associated with spiky search patterns (as shown in figure 5b), which is consistent with empirical measurements of actual activity patterns measured over the Web [35,39]. Therefore, these elements suggest that real life dynamics actually occur with a small propensity of independent search (low $\omega_{ind}$) coupled to strong social influences (high $\omega_{social}$).

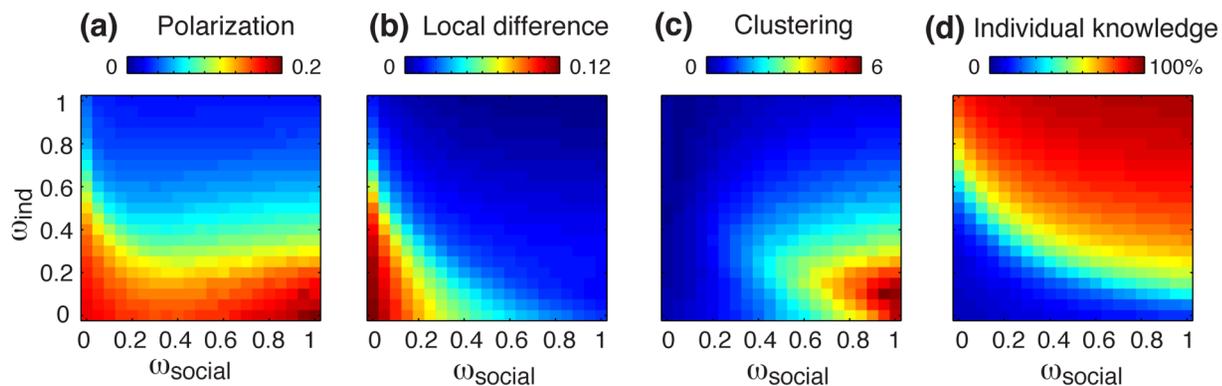

**Figure 4.Collective dynamics of the system as a function of the weight of independent search $\omega_{ind}$, and the weight of social influence $\omega_{social}$.** (**a**) The polarization level indicates how much the views of individuals in the population differ. It is measured as the standard deviation of opinion distribution. A polarization of 0 indicates a global consensus while high values indicate a divergence of opinions in the population. (**b**) The local difference is the average absolute difference between an individuals' opinion and his or her direct neighbors. Low values can indicate a global consensus (such as the example shown Fig. 2c, which lies in the upper right corner of the maps), or local clustering (such as the example show Fig. 2b, which lies in the lower right corner of the map). (**c**) The clustering level is the polarization of the population divided by the local difference. Therefore, the clustering is high when different opinions coexist in the population and a strong agreement is found among neighboring people. (**d**) The average percentage of all available information that are known by individuals. Results are averaged over 50 simulations with model parameters identical to those used in figure 3.
doi:10.1371/journal.pone.0084592.g004

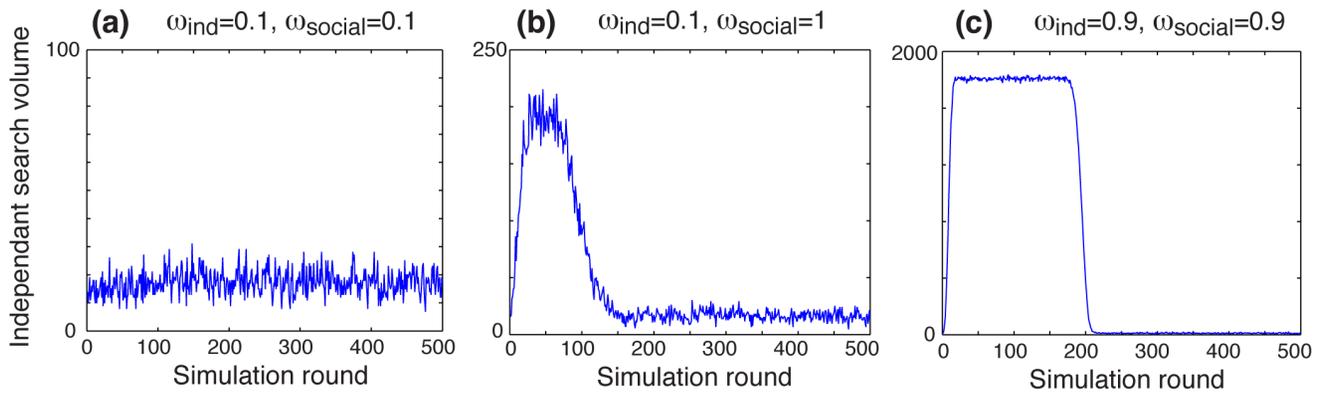

**Figure 5. Three representative examples of the search patterns emerging from the model.** The three examples correspond to the same set of parameters as in Figure 3. (**a**) With low levels of independent search $\omega_{ind}$ and social influence $\omega_{social}$, the search volume is constant and low. (**b**) A spiky search pattern followed by a slow relaxation is visible when $\omega_{ind} = 0.1$ and $\omega_{social} = 1$. (**c**) When both variables are high, the search volume stays high during a certain amount of time, until all individuals become inactive almost simultaneously. The search volume corresponds to the number of individuals who engaged in an independent search per unit of time.
doi:10.1371/journal.pone.0084592.g005

Further simulations of the model in this region of the parameter space (specifically, with $\omega_{ind} = 0.1$ and $\omega_{social} = 0.9$) shade light on how the information flow affects people's risk perception. As illustrated by the example shown in figure 6a, pieces of information tend to spread unequally in the population, where a given item can be intensively exchanged within certain subgroups of people and remain completely ignored by others. In particular, the local flow of information – measured as the number of time an individual $i$ has received a particular piece of information $k$ – exhibits a strongly skewed distribution (figure 6b). These patterns are consistent with the clustering dynamics observed at the population level, as people sharing different subsets of the available information tend develop different risk judgments. The relationship between information flow and risk perception is shown in figure 6c. It appears that individuals expressing extreme opinions are on average less informed than those having a moderate risk judgment. In fact, individuals who take into account a wider diversity of information tend to converge towards a moderate risk judgment. However, the agents in this region of the parameter space are mostly exposed to the opinions of their neighbors and therefore tend to exchange a limited and biased subset of the available information.

## Discussion

In current research, the mechanisms by which people form and revise risk judgments is often investigated at the individual scale, by considering people as isolated units unconnected to their social environment. Existing attempts to describe the *collective* dynamics of risk perception at the population level remain too conceptual to elaborate precise testable predictions. The model that has been introduced in the present work meets the need for quantitative predictions and, therefore, constitutes a testable framework that can help understanding the collective dynamics of the system and complement existing conceptual frameworks well [21].

In addition, the present work contributes to the understanding of collective risk perception in various ways, by (1) showing how clustering and polarization of risk judgment can emerge in a population of interacting agents, (2) identifying parameter space where these phenomena take place, (3) connecting aggregate search patterns, average individual knowledge, and the actual dynamics of risk perception, (4) measuring how conflicting information spread in the population.

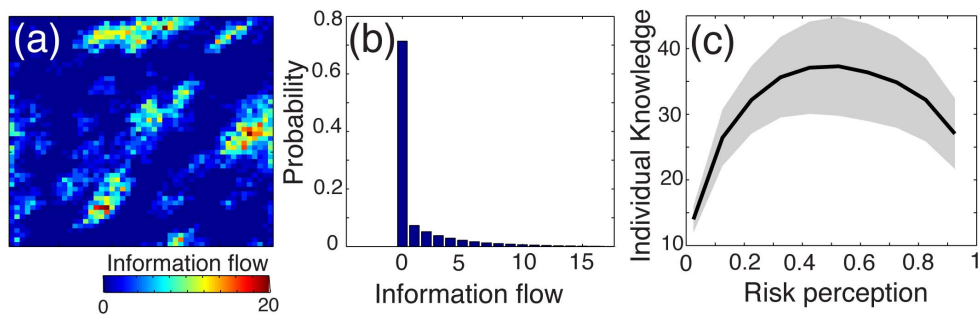

**Figure 6. The dynamics of information flow as observed during simulations.** (**a**) Illustrative example of how one particular piece of information $k$ spreads in the population. The color-coding shows the local information flow, measured as the number of time the information $k$ has been communicated to an individual $i$. Dark blue zones indicate individuals how have never heard of information $k$, whereas those who received the information 20 times are colored in dark red. (**b**) Distribution of the local flow over all pieces of information. The skewed distribution indicates that information spreads unequally in the population. (**c**) The risk perception of individuals as a function of the average number of information they have received. The grey zone indicates the standard deviation of the average. The visible reverse-U shape indicates that individuals expressing extreme opinions are on average less informed than those having a moderate risk judgment. Results are averaged over 50 simulations with parameters $\omega_{ind} = 0.1$ and $\omega_{social} = 0.9$, corresponding to the bottom right corner of the maps presented in figure 4.
doi:10.1371/journal.pone.0084592.g006

In particular, the model highlights that two crucial factors are driving the dynamics of the system: (i) the tendency of individuals to search for their own information in the media environment $\omega_{ind}$, and (ii) the strength of social influence between neighboring people $\omega_{social}$. Different weights given to these two parameters generate a rich variety of collective patterns, such as opinion polarization and opinion clustering. In particular, comparisons with empirical facts suggest that reproducing observed clustering and polarization patterns requires giving a stronger weight to social influence as compared to the role of independent search behaviors. Therefore, an accurate understanding of how people form and revise risk judgments should primarily focus on the nature and frequency of social interactions between people, in contrast to current research trends that mostly consider mass media as the most important source of influence [32]. As a first approximation, the present model assumes the same values of $\omega_{ind}$ and $\omega_{social}$ for all individuals in the population. Nevertheless, one could expect some inter-individual variability on this important behavioral aspect, where some people would tend to give more weight to independent search while others would favor social cues. While the impact of inter-individual variability on the collective outcome has not been studied in the present work, recent research suggests that it could be significant in other social systems [40,41], and therefore should be evaluated in the near future.

Taking some distances from the specific issue of risk perception, the present model relates to other existing research on the emergence of cultural diversity in a population of interacting agents [42,43]. In particular, recent work also came to the conclusion that polarization and diversity of judgments can emerge in a population of people who are exposed to the same set of information, even when assuming different behavioral mechanisms [43–45]. The present model, therefore, complements existing research well, and contribute to the understanding of how diversity of opinions emerge from the combination of local and global influences – considering various mechanisms, social structures and fields of applications.

While the model's predictions can already be explored and compared to empirical data, routes for improvements are numerous. For instance, it remains unclear how the topology of the social network would affect the overall dynamics of the system [46,47]. In fact, most real social networks are scale free networks, with a few individuals being significantly more connected and therefore more influent than others [48]. This aspect of the environment could possibly have an important impact on the system as information may propagate unevenly in the population [49]. Likewise, it is known that people have cultural predispositions to be sensitive to a given risk or not, which may interplay with the formation of their risk judgment [26]. Moreover, neighboring individuals often share similar preferences and behavioral features, which may further enhance the emergence of local basin of agreements [50]. Finally, the model presently assumes a static media environment that remains unchanged over time. In reality, however, media sources of information are themselves subject to the influence of public opinion and other medias [51]. How collective opinion interplays with the structure of the media environment appears as an important question that would require further investigations in the future.

Besides, the model could open interesting applied perspectives, and serve as a prediction tool to help risk assessors anticipating public responses to emerging technologies and innovations. In particular, understanding the precise dynamics that lead to the amplification of risk perception, or reversely to the underestimation of a real danger could facilitate the design and the application of healthcare policies, such as helping doctors to convince a population to adopt certain disease prevention methods, or reversely attenuate people's fears and anxieties towards reasonably safe hazards. This work, therefore, constitutes a starting point that can stimulate an exciting field of research, and lead to concrete predictions of the collective dynamics of risk perception.

## Acknowledgments

The author is grateful to Wolfgang Gaissmaier, Astrid Kause, Jeanne Gouëllo, and Isaac Moussaïd for fruitful discussions and comments.

## Author Contributions

Conceived and designed the experiments: MM. Performed the experiments: MM. Analyzed the data: MM. Contributed reagents/materials/analysis tools: MM. Wrote the paper: MM.